%% file: main.tex
\documentclass[runningheads]{llncs}
\usepackage[T1]{fontenc}
\usepackage{graphicx}
\usepackage{microtype}
\usepackage{makecell}
\usepackage{amsmath}
\usepackage{amssymb}  % assumes amsmath package installed
\usepackage{mathrsfs,color,float}
\usepackage{algorithm, algorithmic, xfrac}
\usepackage{bbding}
\usepackage{tikz}
\usepackage{subcaption}
\usetikzlibrary{fit}
\usetikzlibrary{positioning}

\newcommand{\enve}{\mathcal{E}}
\newcommand{\scriptP}{\mathcal{P}}
\DeclareMathOperator*{\argmax}{arg\,max}
\DeclareMathOperator*{\argmin}{arg\,min}
\begin{document}
\title{Games Over Observation Spaces in Multi-Agent Capture the Flag}
\titlerunning{Games over Observation Spaces}
\author{Mae Frost\Envelope{}\inst{1}\orcidID{0009-0008-9945-979X} \and
Michael Amir\inst{2}\orcidID{0000-0002-5519-0206} \and
Shaunak D. Bopardikar\inst{3}\orcidID{0000-0002-0813-7867}}
\authorrunning{M. Frost et al.}

\newcommand{\MAfix}[1]{%
  {\color{red}\textbf{[MA:} #1\textbf{]}}%
}
\newcommand{\mfNew}[1]{{\color{green}{#1}}}

\institute{Department of Computer Science and Engineering, Michigan State University, East Lansing MI 48824, USA\\ \email{frostric@msu.edu} \and
University of Cambridge, UK\\
\email{ma2151@cst.cam.ac.uk} \and
Department of Electrical and Computer Engineering, Michigan State University, East Lansing MI 48824, USA\\ 
\email{shaunak@msu.edu}}
\maketitle              % typeset the header of the contribution
\begin{abstract}
We consider a multi-agent Capture the Flag (CtF) scenario in a graph-based environment, where a team of attackers seeks to reach designated flag vertices while a defending team attempts to intercept them. In our setting, both teams operate using decentralized heuristic policies.  While the attacking team may choose its heuristic from a diverse library of policies, the defense is restricted to playing a single fixed policy. To overcome this limitation, a centralized defense oracle strategically restricts the portion of the graph visible to each of its agents in order to elicit a wider range of emergent behaviors from its fixed policy. We formalize this interaction as a two-player zero-sum game, where the attacker reasons over its library of heuristics and the defense reasons over the combinatorial space of visibility profiles. To solve this intractably large game, we propose a Double Oracle algorithm to find approximate empirical equilibria, and we empirically validate that observation manipulation can improve the defense's performance.

% \keywords{First keyword  \and Second keyword \and Another keyword.}
\end{abstract}
\section{Introduction}
\label{sec:introduction}
%\MAfix{split this into two paragraphs at least}
Consider an adversarial game of Capture the Flag (CtF) played between two teams of homogeneous agents over a graph. One team, the attackers, tries to reach a set of flag vertices, while the second team, the defenders, tries to prevent flag captures by intercepting the attackers. This leads to a two-team zero-sum game in which the defense minimizes the number of flag captures, while the attackers try to maximize it. In order to accomplish their respective goals, each team uniformly deploys a decentralized policy across its agents, that is, each agent on the same team uses the same policy. In an ideal case, each team would derive an optimal stochastic policy against its opponent, jointly forming a mixed-strategy Nash equilibrium. However, in practice, it is often infeasible to derive optimal policies for either team, especially when the team or graph size is large. 

For this reason, we assume that both teams deploy heuristic policies, which, while suboptimal, are both easier to design and can often be more generalizable across different graphs. With the restriction to decentralized heuristic approaches in place, the question arises of how each team can best shape its behavior to achieve its goal. An intuitive way to do this would be to build a diverse library of heuristics to hopefully allow for a more performant best response to a potential opponent. A second method, however, might be to maintain only a small number of heuristics, and instead shape how each agent perceives the environment by masking portions of the graph. This underlies the central premise of this work, that information restriction can be treated as a strategic resource. While compiling an exhaustive library of heuristics is both labor-intensive and often manual, selectively altering the environment perceived by a fixed policy provides an automated alternative. By selectively withholding specific data, a centralized oracle can effectively restrict each agent's action space to trigger specific heuristic behaviors, such as forcing an agent to hold a choke-point rather than blindly pursuing a target. Under such a profile over the observation and action spaces, each agent on the team potentially plays the game in a different environment, allowing for heterogeneous behaviors despite the homogeneous policy~\cite{amir2026diversity}. In physical deployments, these types of observation restrictions have natural analogs in terms of selective sensor filtering, explicit communication constraints, or dynamic mission assignment. We study the interplay between these two methodologies by considering a \textit{game over visibility profiles} in which the attacking team is allowed access to a library of heuristics, while restricting the defense to only a single policy but allowing it to define an observation mask for each of its agents.

\paragraph{Contribution} The primary contribution of this paper is an algorithm that finds approximate empirical equilibrium policies for both teams in the proposed game over visibility profiles. Our algorithm uses the Double Oracle framework to avoid reasoning over the space of all possible observation masks for each defending agent. While the oracle deciding the attacker's choice of policy is exact, the oracle used to create the visibility profile used by the defense is approximated using deep reinforcement learning techniques to optimize a graph neural network. We then validate our algorithm and oracles in both a small gridworld and a larger setting modeled after urban street maps. Serving as an empirical proof of concept for our framework, these trials demonstrate that manipulating the defense's observation space can impact team behavior enough to shift the value of the CtF game's equilibrium in favor of the defense. Specifically, against our defined set of heuristics, our methodology reduces the expected game value from 1 flag capture to 0 flag captures in the gridworld setting, and from 2 flag captures to 0  flag captures in a 200-vertex urban environment, both representing maximal improvements for the defense. These results provide empirical evidence that observation manipulation can introduce useful heterogeneity to enforce favorable equilibria without engineering new defense policies.

\paragraph{Related Work.} While multi-agent competitive games have been studied in a variety of contexts, our CtF game is most similar to pursuit-evasion~\cite{ma2025multi} and perimeter defense games~\cite{lee2023graph,shishika2020cooperative}, which both feature teams with diametrically opposing goals. However, while the defense must both pursue attackers and protect the flag vertices in CtF, the balance between these goals, particularly across a team, makes defense in CtF an altogether distinct task, a dynamic that equally complicates the attacker's objective. Similar capture the flag settings have also been studied in the context of differential games~\cite{garcia2025capture,wang2023matching}. However, these formulations differ from our own by incorporating a ``safe zone'' to which the attacker must return.  Regardless of the formulation, the large state-space associated with multiple interacting agents has necessitated the creation of algorithmic techniques for solving large games. This led to the creation of the Double Oracle framework~\cite{mcmahan2003planning}, which allows for matrix game solutions to be computed without enumerating either player's action space, as long as a best response oracle is available for both players. The Double Oracle technique has since been extended to a variety of contexts, including extensive form games~\cite{mcaleer2021xdo} and machine learning-derived oracles~\cite{lanctot2017unified}. A final related topic is \emph{agent-environment co-design}, which considers how to optimize not only an agent's behavior for a task, but also the environment around the agent~\cite{gao2025co,li2025scaling,schaff2019jointly,jain2017cooperative}. This idea, that environment design inherently shapes agent behavior, mirrors how our game tries to design optimal CtF graphs for the defending agents. Of particular relevance to our setting are~\cite{gao2025co,li2025scaling}, which utilize a coordinated learning approach to optimize both robot navigation and obstacle placement.

\section{Problem Formulation}
\label{sec:problem-formulation}
In this section, we define the graph-based CtF setting, decentralized heuristics, and the game over visibility profiles. We begin by formulating the game environment.

\subsection{Environment and Agent Formulation}
We formulate the scenario as a discrete time, graph-based Capture the Flag (CtF) game. The game environment is a 7-tuple \(\enve = (G, V_{F}, V_{a}, V_{d}, r_{a}, r_{d}, T)\). \(G = (V,E)\) is a connected, undirected graph, where vertex set \(V\) defines the set of valid locations for agents and edge set \(E\) defines the connectivity between those locations. We define \(V_{F} \subset V\) as a set of static flag locations that the defending team must protect. The initial locations for the defending and attacking teams are given by the sets \(V_{d},V_{a}\subset V\) respectively. We assume the attacker team is of size \(m\) and the defense of size \(n\), such that \(|V_{a}|=m\) and \(|V_{d}|=n\). To ensure non-trivial initial conditions, we impose that team starting locations are strictly disjoint and that no attacker begins at a flag location, i.e., \(V_{d} \cap V_{a} = \emptyset\) and \(V_{F} \cap V_{a} = \emptyset\). 

At any timestep \(t\), the global game state is a tuple \(X(t) = (X_{a}(t) , X_{d}(t))\) and is fully observable to all agents, where \(X_{a}(t) = (x_{a,1}(t), \dots, x_{a,m}(t))\) and \(X_{d}(t) = (x_{d,1}(t), \dots, x_{d,n}(t))\) are tuples representing the current positions (vertices) of the attacking and defending teams respectively. Notably, while we will later introduce restrictions on the state observability for the defense, the attacking team will always maintain access to this global state. As the state transitions, agents either traverse an incident edge to an adjacent location or maintain their current position. At any time step \(t\), each agent still in play is located at exactly one vertex in \(V\), with no restriction on how many agents can be located at any given vertex.

The attacking team \(\mathcal{A}_{a} = \{ a_{1}, \dots, a_{m}\} \) acts according to a decentralized policy \(\pi_{a}\) with the overall goal of \emph{capturing} flags. An attacker agent successfully secures a capture if it navigates to within a capture radius \(r_{a}\in \mathbb{N}_{0}\) (as measured in edges) of any flag vertex. Upon a successful capture, the capturing agent is removed from the environment, while the flag remains active for potential capture by remaining attackers. As the movement of the attacking team is dictated by \(\pi_{a}\), the attackers' objective can be posed as selecting a policy \(\pi_{a}\) that maximizes the total number of flag captures.

The defending team \(\mathcal{A}_{d} = \{ d_{1}, \dots, d_{n}\}\) operates according to a decentralized policy \(\pi_{d}\) with the goal of preventing the attackers from capturing flags. A defender is said to \emph{tag} an attacker if it closes the distance between itself and an attacker to within a tagging radius \(r_{d} \in \mathbb{N}_{0}\). A successful tag results in the removal of both the tagging defender and any tagged attackers from the environment. We give capture priority over tagging should they happen simultaneously. Thus, if an attacker would both capture a flag and be tagged by a defender, the capture will be processed rather than the tag. Furthermore, as tagging is evaluated based only on the final vertex distance after all movements have taken place, agents that `swap' vertices along the same edge will not trigger a tag unless the vertex distance tagging condition holds. Thus, in the case that \(r_{d} = 0\) the defense is incentivized to hold stationary positions rather than pursuing attackers directly. As the defense has directly opposing goals to the attacking team, its objective is to select a policy \(\pi_{d}\) that minimizes the total number of flag captures. 

The game transitions between states in the following manner: all agents select a valid action, i.e., movement to a neighbor or remaining in place, according to \(\pi_{a}\) and \(\pi_{d}\) simultaneously. All actions are then processed at once, updating the agents' locations. Flag captures are then processed, and any attackers that have captured a flag are removed from play. Finally, any agents that meet the tagging condition are removed from play. The game terminates when either one team's agents are entirely removed from play, or when a maximum time horizon \(T\) is reached. Notably, the capture and tagging dynamics described tie the maximum number of flag captures to the size of the attacker team rather than the number of flags. This means that a larger flag set corresponds to a more difficult defensive task, i.e., more locations to defend, as opposed to a change in possible payoffs achievable by either team.

\subsection{Game Utility}
While the environment as described above is fully deterministic, the policies \(\pi_{a}\) and \(\pi_{d}\) may contain stochastic elements. Thus, we will consider the game's utility based on their expected outcome. Let \(C(\pi_{a},\pi_{d} \mid \enve)\) be a random variable representing the total number of flag captures that occur in a single play of the CtF game in environment \(\enve\) with attacker and defender policies \(\pi_{a}\) and \(\pi_{d}\) respectively. We then define the objective function for the game as 
\begin{align}
  J(\pi_{a},\pi_{d}\mid \enve) = \mathbb{E}[C(\pi_{a},\pi_{d} \mid \enve)]. 
\end{align}

Observe that the game utility is strictly zero-sum as the attacking team wishes to maximize \(J(\pi_{a},\pi_{d}\mid \enve)\), while the defending team tries to minimize it.  

\subsection{Policy Structure}
While there exist numerous techniques in the literature that could be adapted to find optimal values of \(\pi_{a}\) and \(\pi_{d}\), we are primarily interested in the theoretical properties of the CtF game when both teams deploy suboptimal, rule-based policies. While such policies may perform worse than what is optimal, they are often significantly easier to design and can often maintain good performance across a wide range of environments. Specifically, we restrict both \(\pi_{a}\) and \(\pi_{d}\) to the class of \emph{decentralized heuristics}, which we define by the following conditions:
\begin{enumerate}
    \item \textbf{Topological Agnosticism:} The policy must be executable across any valid game environment \(\enve\). This ensures the policy produces valid actions regardless of vertex count, edge density, flag placement, or the cardinality of the agent teams.
    \item  \textbf{Uniform Decentralized Execution:} The policy is homogeneous across all agents of a given team. Furthermore, execution is strictly localized; agents independently map their observation of the state to an action without access to any shared hidden state or inter-agent communication.
\end{enumerate}
A trivial example of a decentralized heuristic for the defense might be for each agent to take a step along the shortest path to the nearest attacker and to remain in place if there is no attacker to move towards. This policy is topologically agnostic as it can produce a valid action regardless of the environment configuration or number of agents. Furthermore, it meets the uniform decentralized execution condition as all defending agents use the same process to determine their action, and their action only depends on their observation of the current state of the game. Now that we have formalized the CtF game setting and the policy space, we can turn to our problem formulation. 

\subsection{Games Over Visibility Profiles}
Consider a scenario in which the defending team plays a fixed heuristic policy \(\pi_{d}\). Meanwhile, the attacking team has access to a finite set of heuristic policies \(\Pi_{a} = \{\pi_{a1}, \dots, \pi_{ak}\}\) to choose from. Under these conditions, the objective for the game becomes only a function of the attacker's policy choice. Thus, the attacker may simply evaluate the expected utility of each of its policies against the known defense and select the best response,
\begin{align}
    \pi_{a}^{*} \in \argmax_{\pi_{a} \in \Pi_{a}} J(\pi_{a} \mid \pi_{d}, \enve).
\end{align}
In this formulation, an attacker with access to a diverse set of policies \(\Pi_{a}\) can exploit the static \(\pi_{d}\), leading to poor outcomes for the defense.

In order to restore the defender's agency in this scenario, we must find a way to manipulate the behavior of its agents without altering the underlying policy \(\pi_{d}\). To this end, we introduce the concept of a \textit{visibility profile} over the defending agents. While the defense cannot change the policy \(\pi_{d}\), a centralized oracle can alter the observation space that the defender agents perceive. That is, by restricting the portion of the graph visible to each defender agent at the beginning of the game, the defense can alter the overall behavior of its team. 

Formally, a \emph{visibility profile}, \(p\), over graph \(G\) for a team of \(n\) defenders is defined as an \(n\)-tuple
\begin{align}
    p = ( S_{1} = (V_{1}, E_{1}), \dots ,  S_{n} = (V_{n}, E_{n})),
\end{align}
where \(S_{i}\), \(i=1\dots n\) are induced subgraphs of \(G\). To employ a profile \(p\), the defense restricts the attention of each agent \(d_{i}\) to only the subgraph \(S_{i}\). That is, from the perspective of agent \(d_{i}\) under some profile \(p\), the game environment is \(\enve' = (S_{i}, V_{F}\cap V_{i}, V_{a}\cap V_{i}, V_{d}\cap V_{i}, r_{a}, r_{d}, T)\), restricting the agent's action space accordingly. Similarly, the agent's state observation at time \(t\) becomes \(X'(t) = (X_{a}(t) , X_{d}(t)) \circ V_{i}\), where \(\circ\) denotes the operation of removing all entries in \(X_{a}(t)\) and \(X_{d}(t)\) that do not appear in \(V_{i}\). We emphasize that because the visibility profile is enforced prior to the game's start, defender agents under a profile are unaware of the true environment \(\enve\).

In order to ensure that \(\pi_{d}\) still functions after the profile is applied, we require that each subgraph of the profile forms a valid CtF environment. Specifically, for each subgraph \(S_{i},\, i\in 1\dots n\):
\begin{enumerate}
    \item \(S_{i}\) must be connected;
    \item \(S_{i}\) must contain at least one flag vertex and attacker starting position, \({V_{i} \cap V_{F} \neq \emptyset}\), \(V_{i} \cap V_{a} \neq \emptyset\); and
    \item \(S_{i}\) must contain the starting vertex of agent \(d_{i}\).
\end{enumerate}

Observe that employing a profile can change the behavior of the defense under \(\pi_{d}\). For instance, consider the shortest path heuristic described previously. A carefully designed profile could direct each defender to pursue entirely different targets than in the full-visibility setting. Notably, the profile may restrict the ability of a given defender to see its own team's current locations, which can sever implicit coordination between teammates provided by the heuristic.

Suppose that the defense employs a visibility profile \(p\) over its agents at the start of the game. Since the behavior of the defending team is now determined by both the fixed defense policy \(\pi_{d}\) and the profile \(p\), the objective of the game becomes 
\begin{align}
    J(\pi_{a}, p \mid \pi_{d}, \enve) = \mathbb{E} [C(\pi_{a}, p \mid \pi_{d}, \enve)],
\end{align}
where \(C(\pi_{a}, p |\pi_{d}, \enve)\) is a random variable representing the total number of flag captures that occur in a single instance of CtF in environment \(\enve\) with the attacker playing \(\pi_{a}\) and the defense playing \(\pi_{d}\) under the visibility profile \(p\). This leads to the following minimax optimization problem for a fixed environment \(\enve\):
\begin{align}
    \label{eq:objective}
    \min_{y \in \Delta(\scriptP)} \max_{z \in \Delta(\Pi_{a})} \sum_{p \in \scriptP} \sum_{\pi_{a}\in \Pi_{a}} y(p) J(\pi_{a}, p \mid \pi_{d}, \enve) z(\pi_{a}),
\end{align}
where \(\Delta(\scriptP)\) and \(\Delta(\Pi_{a})\)  are the probability simplices over the space of valid visibility profiles and attacker heuristics, respectively. We impose that both teams' choices of strategy are made simultaneously; neither team observes the opponent's strategy distribution or realized choice prior to execution. The solution to this objective represents the Nash equilibrium of a CtF game where the defense reasons over the space of visibility profiles while the attackers reason over a set of heuristics. For notational clarity, we will refer to possible values of \(y\) and \(z\) as the defender and attacker \emph{strategies}, as to distinguish them from the CtF \emph{policies} \(\pi_{d}\) and \(\pi_{a}\in \Pi_{a}\).

\textbf{Problem Statement:} Determine a methodology for solving the objective in Equation~\eqref{eq:objective} by finding empirical equilibrium strategies \((y^{*}, z^{*})\).

\section{Methodology}
\label{sec:methods}
In this section, we describe a Double Oracle approach for solving the zero-sum visibility profile game described above. While zero-sum games can typically be solved efficiently through the use of linear programming, the large quantity of pure strategies available to the defense makes this challenging in practice. Since each pure strategy for the defense corresponds to an \(n\)-tuple of subgraphs of the game graph \(G\), we have that the total number of possible defense strategies is upper bounded by \(|\scriptP| < 2^{n|V|}\), where \(n\) is the number of defenders. To avoid enumerating this strategy space, we utilize the Double Oracle technique which allows us to only explore relevant portions of the visibility profile space. 

The formal statement of our algorithm is given in Algorithm~\ref{alg:DO} and is summarized as follows. The set \(Y\) contains the defense visibility profiles that have been generated so far, while \(Z\) is the current subset of \(\Pi_{a}\) under consideration. We initialize \(Y\) to contain the visibility profile corresponding to full visibility for all agents. This ensures that the equilibrium value for the defense is no worse than the full-visibility baseline. \(Z\) is initialized with a randomly selected attacker policy. In each iteration of the algorithm, the profiles and policies in \(Y\) and \(Z\) form a two-player zero-sum \emph{sub-game} where \(Y\) and \(Z\) serve as the sets of pure strategies available to each player. \(U\in \mathbb{R}^{|Y|\times |\Pi_{a}|}\) is a utility matrix which holds the estimated expected game payoff for the current set of visibility profiles against \emph{all} attacker policies. The subroutine \(Nash(Y,Z,U_{k})\) finds equilibrium strategies, \(y_{k} \in \Delta(Y)\) and \(z_{k} \in \Delta(Z)\), for the current sub-game using a standard linear programming formulation, drawing the game's utilities from the matrix \(U_{k}\). The \emph{Profile Oracle} then calculates a visibility profile \(p_{br} \in \scriptP\) that is a best response to the current attacker mixed policy \(z_{k}\). Similarly, the \emph{Attacker Oracle} determines a pure policy \(\pi_{br} \in \Pi_{a}\) as the best response to the current defender mixed policy \(y_{k}\). Note that \(p_{br}\) and \(\pi_{br}\) are drawn from the full sets \(\scriptP\) and \(\Pi_{a}\), not just from \(Y\) and \(Z\). The best response profile and policy are then added to the sets \(Y\) and \(Z\) respectively, and a new row is computed for the utility matrix \(U\) if \(p_{br}\) is a new addition to \(Y\).

\begin{algorithm}[htbp]
    \caption{Double Oracle for Visibility Profile Game}
    \begin{algorithmic}[1]
    \label{alg:DO}
    \STATE Initialize \(Y\) with profile \(p_{0} = (S_{1} = G, \dots, S_{n}=G)\)
    \STATE Initialize \(Z\) with a randomly selected policy from \(\Pi_{a}\)
    \STATE Initialize \(U\) by estimating \(\hat{J}(\pi_{a},p_{0}\mid \pi_{d}, \enve)\,\, \forall \pi_{a} \in \Pi_{a}\)
    \STATE \(k \leftarrow 0\)
    \WHILE{not converged}
        \STATE \(U_{k} \leftarrow U\)
        \STATE \((y_{k}, z_{k}) \leftarrow Nash(Y,Z,U_{k})\)
        \STATE \(p_{br} \leftarrow ProfileOracle(z_{k})\)
        \STATE \(\pi_{br} \leftarrow AttackerOracle(y_{k},U_{k})\)
        \STATE \(Y \leftarrow Y \cup \{p_{br}\} \)
        \STATE \(Z \leftarrow Z \cup \{\pi_{br}\}\)
        \STATE Update \(U\) for profile \(p_{br}\) if needed. 
        \STATE \(k \leftarrow k+1\)
    \ENDWHILE
    \end{algorithmic}
\end{algorithm}

The algorithm formally converges if, in some iteration, both \(p_{br}\) and \(\pi_{br}\) are already in \(Y\) and \(Z\) respectively. That is, the algorithm terminates if both oracles cannot find a best response that is not already part of the sub-game. More commonly, however, Double Oracle algorithms are set to terminate if the difference in utility between the sub-game's Nash equilibrium on subsequent iterations falls below some threshold \(\delta\)~\cite{mcmahan2003planning}. The maintenance of the utility matrix \(U\) is borrowed from the Policy Space Response Oracle (PSRO) framework~\cite{lanctot2017unified} and is not strictly necessary. However, it can greatly speed up computation through memoization as utilities are calculated by policy rollout. Under the assumption that both oracles can accurately give the best response, the double oracle method is known to converge in a finite number of iterations~\cite{mcmahan2003planning}. However, as our Profile Oracle is approximated, Algorithm~\ref{alg:DO} terminates in an approximate empirical equilibrium.  We now cover the details of each of the oracles used in Algorithm~\ref{alg:DO}.

\subsection{Attacker Oracle}
Recall that the purpose of the Attacker Oracle is to quickly choose a best response pure policy \(\pi_{br}\in \Pi_{a}\) to the current profile mixed strategy \(y_{k}\). In practice, the set of pure policies available to the Attacker is significantly smaller than the space of pure profiles, that is, \(|\Pi_{a}| \ll |\scriptP|\). This allows us to calculate the best response for the attacker directly as,
\begin{align}
    \label{eq:attacker-br}
    \pi_{br} \in \argmax_{\pi \in \Pi_{a}} \,[y_{k}' U_{k}]_{\pi},
\end{align}
where \(y_{k}\in \mathbb{R}^{|Y|\times 1}\) is the defender mixed strategy over the visibility profiles within the sub-game, and \(U_{k}\) is the current utility matrix.

\subsection{Profile Oracle}
Unlike the Attacker Oracle, the Profile Oracle must select a pure strategy from the combinatorial space of all valid visibility profiles, \(\scriptP\). Formally, the oracle's task is to select visibility profile \(p_{br} \in \scriptP\) that minimizes the expected CtF game utility against the current mixed attacker strategy, \(z_{k} \in \Delta(Z)\), that is,
\begin{align}
    \label{eq:part-oracle-obj}
    p_{br} \in \argmin_{p\in \scriptP} \sum_{\pi_{a} \in Z} z_{k}(\pi_{a}) J(\pi_{a}, p \mid \pi_{d}, \enve).
\end{align}

As \(|\scriptP|\) scales exponentially with the size of the game graph \(G\), exhaustive evaluation of Equation~\ref{eq:part-oracle-obj} is intractable for all except the smallest of graphs. For this reason, we instead approximate the Profile Oracle using Deep Reinforcement Learning (DRL). This is accomplished by reframing the selection of the profile as an iterative vertex selection task, where a centralized \emph{visibility agent} sequentially chooses vertices to exclude from each defender's observation set to support the performance of \(\pi_{d}\). 

This process of sequential vertex exclusion is modeled as a Markov Decision Process (MDP) defined by the tuple \((\mathcal{S}, A, r, \mathcal{T} )\). The state at time step \(t\), \(s_{t}\in \mathcal{S}\), encodes the graph topology of the CtF game \(G\), the flag locations, agent starting positions, and the current state of the constructed profile. The action space \(A\) is composed of \(n\cdot|V|\) discrete actions, corresponding to removing a specific vertex from the observable subgraph of a specific defender agent. The action space also includes a single designated ``commit'' action, which causes the end of the episode. The transition function, \(\mathcal{T}\), is fully deterministic, simply updating the state to reflect the updated visibility profile based on the removal of the chosen vertex. The reward function, \(r\), is sparse, only evaluated after the agent has chosen the commit action finalizing the choice of profile, \(p\). The reward is determined by using the negative number of flag captures that occur in a simulated CtF game, using the profile \(p\) and an attacker policy \(\pi_{a} \sim z_{k}\).

To solve this MDP, we parameterize the visibility agent as a neural policy \(\pi_{\Theta}(a_{t} \in A| s_{t} \in \mathcal{S})\), optimizing the parameters \(\Theta\) using an Actor-Critic architecture. Here, we utilize a Graph Neural Network (GNN) to parametrize our policy due to the graph-based state space of the MDP. However, in practice the technique is agnostic to the choice of function approximator. We now turn to a discussion of some modifications we apply to the profile MDP to better facilitate the optimization of the model parameters, \(\Theta\).

\subsubsection{Reward Shaping}
As mentioned above, the reward associated with a generated visibility profile is determined by using the negative number of flag captures that result from simulating the CtF game. In practice, this means that the only possible rewards that can be obtained for a given CtF environment \(\enve\) are integers from the set \(\{-|V_{F}|, \dots , 0\}\). While theoretically grounded in optimizing the objective given in Equation~\eqref{eq:part-oracle-obj}, the sparse reward signals can often lead to high variance gradient updates during the optimization of \(\pi_{\Theta}\), particularly in early training iterations. In preliminary evaluations, we observed that optimizing the visibility agent using only the flag captures frequently struggled early on. To help stabilize the training process in early iterations, we augment the objective with a surrogate reward, \(r_{dense}\), that encourages the visibility agent to generate profiles that minimize the shortest path distance between defender and attacker agents at the termination of the CtF game. More formally, let \(\mathcal{A}_{a}^{f}, \mathcal{A}_{d}^{f} \subset V\) be the locations of any active attacker and defender agents respectively at the conclusion of a CtF game before the final checks for capture and tagging, the dense reward when \(|\mathcal{A}_{a}^{f}|,\,|\mathcal{A}_{d}^{f}| >0\) is then given by,
\begin{align}
    r_{dense}(p) = -\frac{ \min \{ dist(a^{f}, d^{f}) \mid a^{f}\in \mathcal{A}_{a}^{f},\,  d^{f}\in \mathcal{A}_{d}^{f} \} }{\max \{ dist(i,j) \mid i,j\in V \}},
\end{align}
where \(dist: V \times V \rightarrow \mathbb{N}\) gives the shortest path distance between vertices in \(G\), and 0 otherwise. The total reward observed by the visibility agent is given by 
\begin{align}
    r_{agent} = r(p) + \beta \cdot  r_{dense}(p),
\end{align}
where \(\beta\) is a dynamic weighting parameter, controlling the influence of the surrogate reward. At the beginning of training, \(\beta\) is initialized to 1, providing an additional reward signal to help guide the agent in early iterations.  However, to ensure that the visibility agent ultimately optimizes Equation~\eqref{eq:part-oracle-obj} without becoming overly biased by \(r_{dense}\), \(\beta\) is linearly annealed to 0 over the course of policy optimization. 

\subsubsection{Profile Action Masking}
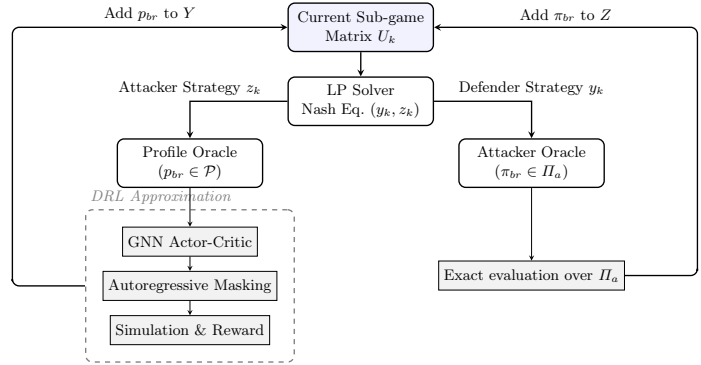
\begin{figure}[htbp]
    \centering
    \resizebox{0.75\textwidth}{!}{\input{tikz-images/algo-diagram}}
    \caption{A block diagram visualizing the major components and information flow of Algorithm~\ref{alg:DO}.}
    \label{fig:alg-diagram}
\end{figure}

A critical challenge in generating the profile \(p\) is ensuring that the final subgraphs each form a valid CtF environment. While this could, in theory, be enforced by introducing additional penalties in the reward function, such a scheme would likely slow down the optimization process considerably. Instead, we employ autoregressive action masking during the policy's forward pass to enforce the following conditions. Suppose an action at time \(t\), \(a_{t}\), would cause the removal of some vertex \(j\) from subgraph \(S_{i}\) creating the updated subgraph \(S' = (V', E')\) for some defender \(i\), we require:
\begin{enumerate}
    \item The vertex \(j\) has not already been removed from \(S_{i}\), i.e., \(j \in V_{i}\);
    \item \(S'\) is connected;
    \item \(S'\) contains at least one flag, i.e., \(V_{F} \cap V' \neq \emptyset\);
    \item \(S'\) contains at least one attacker starting vertex, i.e., \(V_{a} \cap V' \neq \emptyset\); and
    \item \(S'\) contains the starting location of defender \(i\).
\end{enumerate}
At each time step \(t\), we compute a binary mask to determine the legal actions given the current topology of the profile. This mask is then used to set the logits corresponding to invalid vertices to \(-\infty\), ensuring that these vertices will never be selected and that the resulting policy \(\pi_{\Theta}(a_{t}| s_{t})\) can only sample valid profiles. On a practical note, enforcing the second condition (\(S'\) is connected) is a non-trivial operation, requiring computation and masking of the articulation points of \(S_{i}\). In general, this can be done in time \(\mathcal{O}(|V_{i}| + |E_{i}|)\) using the method of Hopcroft and Tarjan~\cite{hopcroft1973algorithm}. However, since this process must be repeated for all \(n\) defenders, it can become burdensome when either the graph or team size is particularly large. In such cases, it may prove more efficient to employ a reward shaping approach to maintaining connectivity, rather than action masking. A complete schematic of Algorithm~\ref{alg:DO}'s execution loop is shown in Figure~\ref{fig:alg-diagram}.

\subsection{On Convergence and Termination}
The standard Double Oracle framework has been shown to converge to an exact Nash equilibrium provided that the oracles return a true best response. While this is the case for our Attacker Oracle, our Profile Oracle is approximated via Deep Reinforcement Learning (DRL), causing our approach to fall under the framework of PSRO~\cite{lanctot2017unified}. Under PSRO, if the approximate oracle is assumed to find an \(\epsilon\)-best response, then the overall solution can be shown to be an \(\epsilon\)-Nash equilibrium~\cite{mcmahan2003planning,mcaleer2021xdo}. However, as the DRL-based oracle lacks a formal \(\epsilon\)-best response bound, the outputs of Algorithm~\ref{alg:DO} form an \emph{empirical equilibrium} rather than an exact Nash equilibrium. 

For this reason, we employ a termination condition based on the stagnation of the sub-game equilibrium value. Let \(v_{k}\) denote the value of the sub-game equilibrium in the \(k\)-th iteration of Algorithm~\ref{alg:DO}. The algorithm terminates when the addition of a new profile policy fails to reduce the equilibrium value by more than a hyperparameter threshold \(\delta\). Specifically, we terminate if \(|v_{k} - v_{k-1}| < \delta\) \emph{and} no new attacker heuristic has been added to the sub-game. This indicates that the Profile Oracle has plateaued in its ability to find new visibility profiles that can improve performance against the current attacker mixed strategy. 

 We know that in any iteration of Algorithm~\ref{alg:DO}, the Attacker Oracle produces a true best response as it performs an exact argmax over its finite policy library. However, due to the combinatorial complexity of finding an optimal profile, it is difficult to measure the quality of the approximate Profile Oracle. However, because the value of the CtF game is strictly lower bounded by 0, we have that any strategy returned by the Profile Oracle with a value of 0 must be a true best response.  Thus, under the assumption that the utility matrix \(U\) accurately estimates the expected game payoffs, the special case of convergence to a zero-capture final solution guarantees that the resulting strategies form a true Nash equilibrium.

\section{Experiments}
\label{sec:experiments}
In this section, we empirically evaluate the performance of Algorithm~\ref{alg:DO} and the approximate Profile Oracle. We begin with an illustrative example played over a 12-vertex gridworld, which allows for easy qualitative analysis of the profile. We then gain insight into the average performance of Algorithm~\ref{alg:DO} through a set of random trials on larger graphs with more realistic topologies. We then conclude with a detailed case study of the algorithm's execution in one of these randomized trials. All experiments are performed on a computer with an Intel Core i7-12700 CPU, a GeForce RTX 3060 Ti, and 32 GB of memory.\footnote[1]{Code available at https://github.com/rfrost114/Visibility-Profile-Games}

\subsection{Experimental Setup}
We evaluate our approach under two classes of CtF graphs and across a small set of attacker and defender heuristics.

\begin{figure}[htbp]
    \centering
        \begin{subfigure}[b]{0.45\textwidth}
        \centering

        \includegraphics[width=\textwidth]{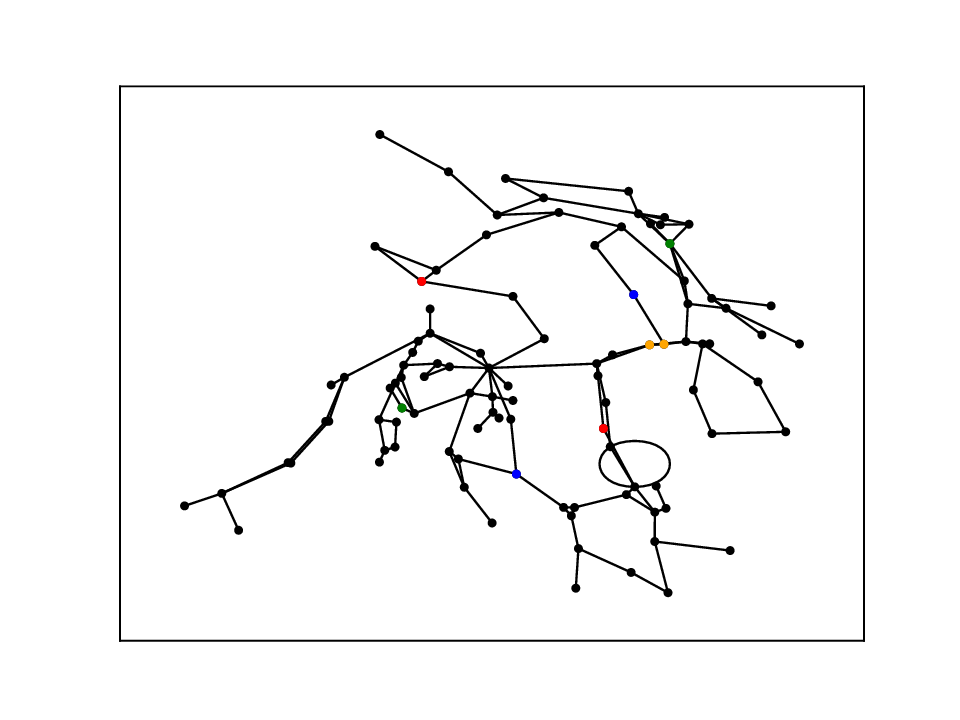}
        \caption{100-Vertex Environment}
        \label{fig:osm100}
    \end{subfigure}
    \hfill 
    \begin{subfigure}[b]{0.45\textwidth}
        \centering
        \includegraphics[width=\textwidth]{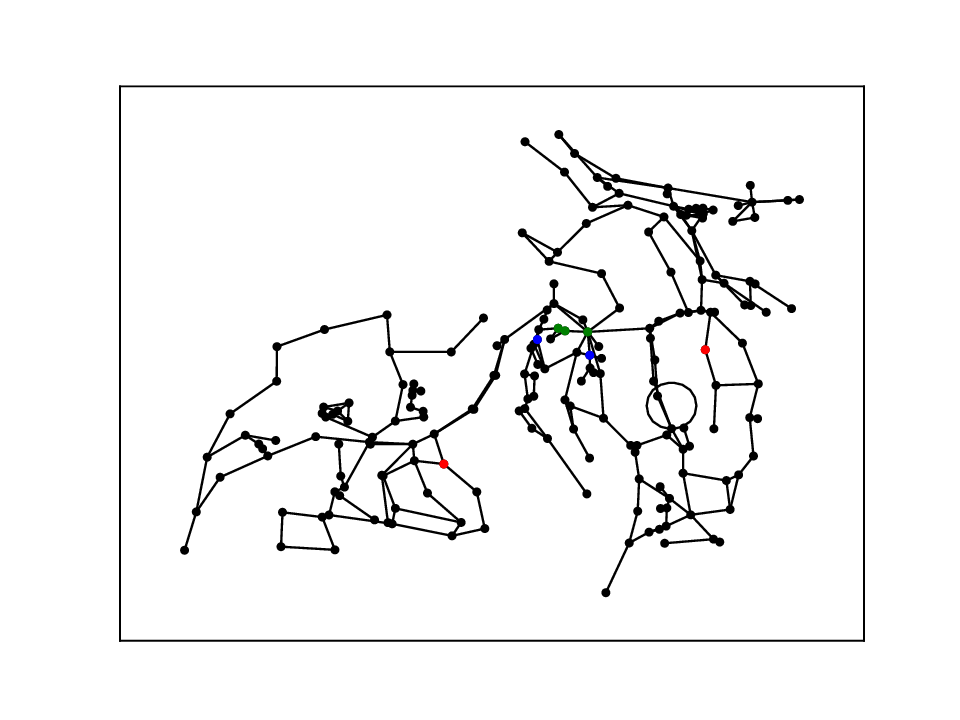}
        \caption{200-Vertex Environment}
        \label{fig:osm200}
    \end{subfigure}

    \caption{One configuration of the OSM-based CtF environments used in the randomized trials. Flags are shown in green, attacker starting locations in red, and defender starting locations in blue.}
    \label{fig:osm}
\end{figure}

\paragraph{Graph Settings}
As noted previously, the first experiment demonstrates our approach's performance on a 12-vertex gridworld. The scenario features one defending agent, one attacking agent, and one flag, each positioned in one of the ``corner'' vertices. The defending team is given a tagging radius of \(r_{d} = 1\), while the attackers are given a smaller capture radius of \(r_{a} = 0\). The time horizon in this setting is set to \(T=50\). The setting for both the randomized trials and the final case study are played over two graphs extracted from OpenStreetMap (OSM) data~\cite{haklay2008openstreetmap}, with 100 and 200 vertices respectively. For both of these graphs, the scenario is a two-versus-two setting featuring 2 attackers and 2 defenders. The 100-vertex graph features 2 flag vertices that are spread across the graph. In contrast, the 200-vertex graph has 3 tightly clustered flag vertices. The OSM graphs are shown in Figure~\ref{fig:osm}. In this setting, both the capture and tagging radius are set to \(r_{a} = r_{d} = 1\), and games are played for a maximum of \(T=200\) steps.

\paragraph{Heuristic Library}
For our experiments, we utilize a set of baseline decentralized  heuristics for both the attacker and defender. In each game, the attacker oracle uses a library of three distinct routing behaviors, while the defense is locked to one of two preset policies. These policies are summarized in Table~\ref{tab:heuristics}. Notably, while each of the attacker heuristics contain stochastic elements, they only arise when the heuristic must break a tie between two equal options. In general, however, we observe that the behavior of the heuristics is largely deterministic within our experimental settings.

\subsection{Implementation Details}
To solve the profile MDP, we parameterize the profile policy using a GNN. The GNN is then optimized using Maskable Proximal Policy Optimization (Masked PPO)~\cite{huang2022mask}. We do not use a warm starting mechanism in our experiments, with the GNN being retrained from scratch in each iteration. As mentioned above, the state representation that the GNN receives as input is in two parts: a CtF graph representation (which remains the same across all states) and an encoding of the current profile (which changes between states). The CtF graph representation is given by a vertex feature matrix, where each vertex is represented by a 21-dimensional feature vector. This feature vector can be broken into three conceptual sections: a 4-dimensional one-hot encoding of the vertex's presence in \(V_{F}\), \(V_{a}\), or \(V_{d}\); a 4-dimensional encoding of local graph information (the vertex's degree along with the minimum, maximum and mean degree of its neighbors); and a 13-dimensional encoding of the vertex's global position within the graph using 8 Laplacian eigenvectors~\cite{dwivedi2023lap-eig} and 5 random walk~\cite{dwivedi2021walkUP} positional embeddings. The current profile is represented by a \(|V|\times n\) binary matrix, where the entry at row \(i\) and column \(j\) is 1 if vertex \(i\) is included in the current profile for defender \(j\) and 0 otherwise. These two matrices are then concatenated to form the final state representation matrix: \(s_{t} \in \mathbb{R}^{|V|\times (21+n)}\). 

\begin{table}[htbp]
\caption{Agent Heuristics}
\label{tab:heuristics}
    \centering
    \begin{tabular}{p{0.15\linewidth} p{0.25\linewidth} p{0.5\linewidth}}
        \textbf{Role} & \textbf{Heuristic Name} & \textbf{Behavior Description} \\ \hline
         Defender & Baseline  & Navigates along the shortest path to the nearest attacker.\\ \hline
         Defender & Patrolling & Patrols between high degree vertices in the vicinity of the flags. \\ \hline
         Attacker & Shortest Path  &  Navigates along the shortest path to the nearest flag\\ \hline
         Attacker & Evasive Routing & Navigates to the nearest flag while maintaining distance from nearby defenders. \\ \hline
         Attacker & Receding Horizon & Assigns a vertex score within a \(5\)-hop neighborhood based on flag and defender proximity, and steps based on the least score path.\\ \hline
    \end{tabular}
\end{table}

The message-passing architecture in our GNN is composed of two Graph Attention Network (GAT) layers~\cite{velivckovic2018gat} each with a hidden dimension of 64. The first layer utilizes 4 attention heads, each of which generates a 16-dimensional embedding. The second layer then uses a single attention head to aggregate these embeddings into a single 64-dimensional embedding for each vertex. After each layer, the ELU activation function is used to provide nonlinearity. To help prevent the oversquashing phenomenon that can plague shallow GNNs, we utilize a global virtual vertex within our network which helps to aggregate information between distant vertices in the graph~\cite{pham2017graph,southern2025understanding}. To help stabilize policy optimization and reduce variance in the critic's advantage estimates, we employ an asymmetric, centralized critic architecture during training~\cite{pinto2017asymmetric}. While the actor model is constrained to the state space described above to ensure it learns a generalized best response, we allow the critic model privileged access to the specific attacker heuristic sampled for the current episode. As the critic model is only utilized in training to guide the actor's gradient updates and is entirely unused during evaluation, this methodology maintains the information boundaries needed to ensure a generalized Profile Oracle while increasing the training efficiency. 

During policy optimization, the network was trained for a maximum of one million time steps, incorporating an early stopping mechanism to halt training on convergence. This mechanism halts training early if the average reward across 100 episodes rises above -0.15.  In each update cycle, we collected 2000 environment steps per rollout and updated the network using a batch size of 256 and a learning rate of \(3\times10^{-4}\). To stabilize training and encourage exploration, we used a PPO clip range of 0.15, a target KL divergence of 0.015, and an entropy coefficient of 0.01. As the profile generation process is a finite-horizon task with rewards only evaluated at termination, we used a discount factor and Generalized Advantage Estimation (GAE) lambda of 1. This helps to ensure that early node-masking actions are not arbitrarily penalized relative to actions taken immediately prior to episode termination.

Finally, the simulation required throughout Algorithm~\ref{alg:DO} was implemented using the GAMMS Python package~\cite{patil2026gamms}, which allows for the efficient implementation of multi-agent games in graph settings. Due to the largely deterministic behavior of our test heuristics, simulations for calculating values in the utility matrix \(U_{k}\) are run 20 times, with the mean flag captures reported.

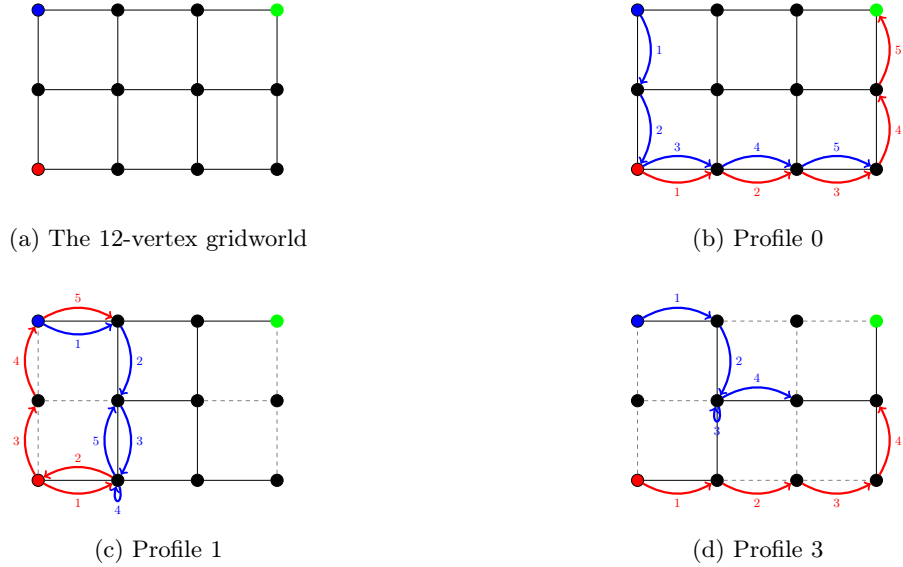
\begin{figure}[htbp]
    \centering
    
    % --- Row 1 ---
    \begin{subfigure}[b]{0.35\textwidth}
        \centering
        \resizebox{\textwidth}{!}{\input{tikz-images/gridworld}}
        \caption{The 12-vertex gridworld}
        \label{fig:base-gw}
    \end{subfigure}
    \hfill 
    \begin{subfigure}[b]{0.35\textwidth}
        \centering
        \resizebox{\textwidth}{!}{\input{tikz-images/gw-base}}
        \caption{Profile 0}
        \label{fig:gw-base}
    \end{subfigure}
    
    \vspace{1em} 
    
    % --- Row 2 ---
    \begin{subfigure}[b]{0.35\textwidth}
        \centering
        \resizebox{\textwidth}{!}{\input{tikz-images/gw-beats-ret}}
        \caption{Profile 1}
        \label{fig:gw-beat-ret}
    \end{subfigure}
    \hfill 
    \begin{subfigure}[b]{0.35\textwidth}
        \centering
        \resizebox{\textwidth}{!}{\input{tikz-images/gw-final}}
        \caption{Profile 3}
        \label{fig:gw-final}
    \end{subfigure}
    \caption{Evolution of the defense profile over multiple double oracle iterations. (a) The base gridworld with the flag (green), attacker starting location (red), and defender starting location (blue) shown. (b) When the defender uses no profile, the Evasive Routing attacker (red arrows) captures the flag. (c) The Profile Oracle drops 2 vertices (gray) and 6 incident edges (dashed) allowing the defender (blue arrows) to tag the attacker. (d) The final profile allows the defense to tag the attacker regardless of heuristic.}
    \label{fig:gridworld_trap}
\end{figure}

\subsection{Gridworld Experiment}
The first experiment takes place in the 12-vertex gridworld graph shown in Figure~\ref{fig:base-gw}. The defense utilizes the Baseline heuristic, while the attacker has access to all of the attack heuristics listed in Table~\ref{tab:heuristics}. As a baseline, when the defense plays in a full visibility setting, the Shortest Path and Evasive Routing have an expected payoff of 1 flag capture, while the Receding Horizon policy has an expected payoff of 0. All three policies deterministically give these performance levels in the gridworld. Under this baseline, the attacker will choose to use either the Shortest Path or Evasive Routing heuristic, achieving a payoff of 1 for the CtF game. Algorithm~\ref{alg:DO} will allow the defense to outperform this baseline. 

In the experiment, the attacker sub-game is initialized with the Evasive Routing heuristic, while the defense begins by using no visibility profile (Profile 0), leading to a sub-game equilibrium payoff of 1 capture. The unrolling of these policy choices is shown in Figure~\ref{fig:gw-base}. The Profile Oracle then calculates a best response profile (Profile 1) to the Evasive Routing attacker, and the Attacker Oracle adds the Shortest Path Heuristic to the sub-game. The best response profile effectively mitigates the Evasive Routing attacker leading to a payoff of 0 captures (Figure~\ref{fig:gw-beat-ret}). However, the Shortest Path heuristic added to the sub-game is unaffected by this new defender behavior and can still score the flag capture. This leads to the sub-game equilibrium value at the beginning of the algorithm's second iteration remaining at 1 flag capture, with the defender mixing between its two profiles and the attacker always playing the Shortest Path heuristic.

The algorithm iterates several more times, with the defense next creating a profile (Profile 2) that beats the Shortest Path heuristic but not the Evasive Routing heuristic, leading to mixed strategies for both players and an expected equilibrium value of \(0.5\). Eventually, at termination, the defense has access to two separate profiles (Profiles 3 and 4), both of which can mitigate the Evasive Routing and Shortest Path heuristics, over which the defense mixes equally. The Receding Horizon heuristic ends up always being suboptimal and is never added to the sub-game, leading to a final attacker strategy of mixing evenly over Evasive Routing and Shortest Path. The final sub-game equilibrium value is 0 captures, which corresponds to the maximum possible improvement for the defense. The unrolling of the game when the defense applies Profile 3 against the Shortest Path Attacker is shown in Figure~\ref{fig:gw-final}. Interestingly, both Profiles 1 and 3 learn to exploit a fail case in the defense policy, in which the defender holds position if it cannot see the attacker. By strategically hiding the attacker's position at a critical time-step, the defense can better position itself and tag the attacker.

\subsection{Randomized Trials}
In our second experiment, we utilize the two OSM graphs described above to test the performance of our technique on arbitrary starting conditions. For each graph we generate 10 random configurations of agent starting positions and flag locations. We then report the mean performance of Algorithm~\ref{alg:DO} over all configurations. For the 100-vertex OSM graph, we report results for the both the Baseline and Patrolling defender heuristics. For the 200-vertex OSM graph we report only the Patrolling defender's results for brevity. As a baseline, we report the average game payoff when the defense uses a full visibility profile averaged over 100 simulations per configuration. As a secondary baseline, we also generate a set of 100 random valid visibility profiles for each configuration and report the mean CtF payoff when the attacker chooses its best response policy over 100 rollouts. While we continue to use 20 policy rollouts for calculating \(U_{k}\) during execution, we reevaluate the utility for any profile that appears in the equilibrium on termination using 100 rollouts to ensure a greater degree of accuracy. The results of these trials can be seen in Table~\ref{tab:results}.

\begin{table}[htbp]
\caption{Randomized Trial Results: Mean over 10 Starting Configurations}
\label{tab:results}
    \centering
    \begin{tabular}{p{0.15\linewidth} p{0.15\linewidth} p{0.2\linewidth} p{0.2\linewidth} p{0.2\linewidth}}
        \textbf{Heuristic} & \textbf{Graph} & \makecell[l]{\textbf{Full} \\\textbf{Visibility}} & \makecell[l]{\textbf{Mean} \\\textbf{Random}} & \textbf{Algorithm~\ref{alg:DO}}\\ \hline
       Baseline & 100-Vertex & 0.80\(\pm0.40\) & 0.97\(\pm0.59\) & 0.50\(\pm0.50\) \\ \hline
       Patrolling & 100-Vertex & 1.30\(\pm0.64\) & 1.28\(\pm0.64\) & 0.60\(\pm0.48\) \\ \hline
       Patrolling  & 200-Vertex & 1.40\(\pm0.66\) & 1.49\(\pm0.59\)& 0.65\(\pm0.63\) \\ \hline
    \end{tabular}
\end{table}

Across both heuristics and both graphs, the mean game payoff for the defense improves when utilizing Algorithm~\ref{alg:DO} as compared to using full visibility. Furthermore, using a random choice of profile is, on average, disadvantageous compared to simply using full visibility. We also observe that for some trials, Algorithm~\ref{alg:DO} cannot find a profiling strategy that improves performance. In some cases, this was because the initial conditions were so favorable to one team that the baseline payoff could not mathematically be improved. In other cases, it is less clear if no such profile exists or if the Profile Oracle simply did not find it. We can gain some insight into these cases by comparing the final payoffs per configuration in the 100-vertex graph for the two separate defense heuristics. Save for one configuration, the policies achieve the same final payoffs after applying Algorithm~\ref{alg:DO}. This suggests that the final payoffs obtained in these trials are at least a local minimum in the space of profile performances. The final observation we draw is that, while the Baseline heuristic outperforms the Patrolling heuristic significantly in the full visibility setting, the two heuristics have similar expected performance after applying Algorithm~\ref{alg:DO}. This suggests, the Patrolling heuristic gains more from the heterogeneity introduced through observation profiling than the Baseline heuristic.  

\subsection{200-Vertex Case Study}
\begin{figure}[htbp]
    \centering
        \begin{subfigure}[b]{0.45\textwidth}
        \centering

        \includegraphics[width=\textwidth]{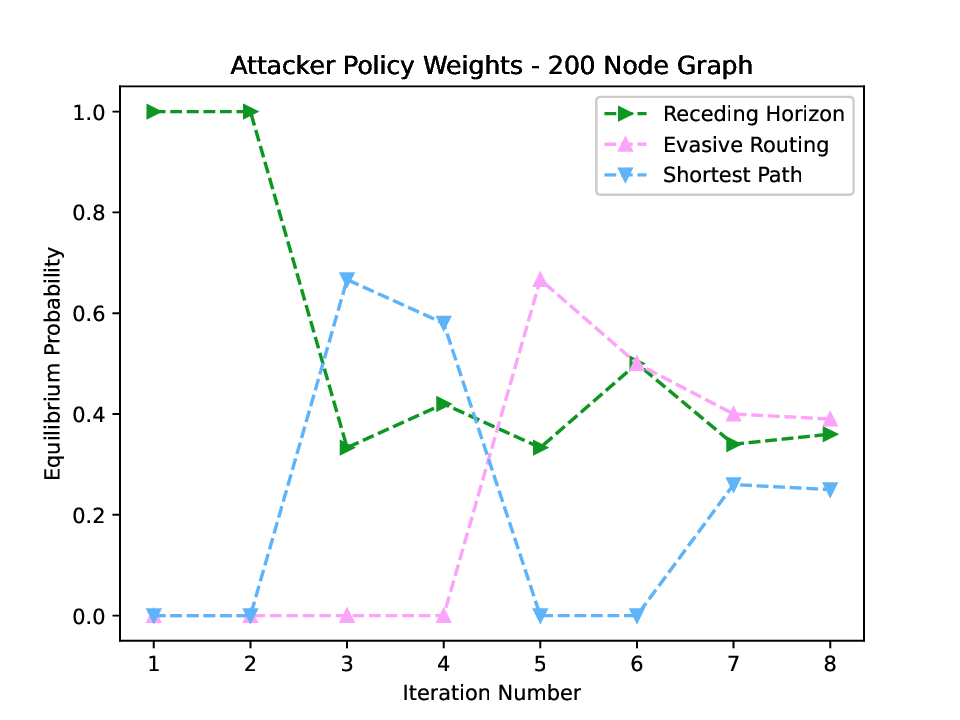}
        \caption{Sub-game Attacker Mixed Strategies}
        \label{fig:200-mixed}
    \end{subfigure}
    \hfill 
    \begin{subfigure}[b]{0.45\textwidth}
        \centering
        \includegraphics[width=\textwidth]{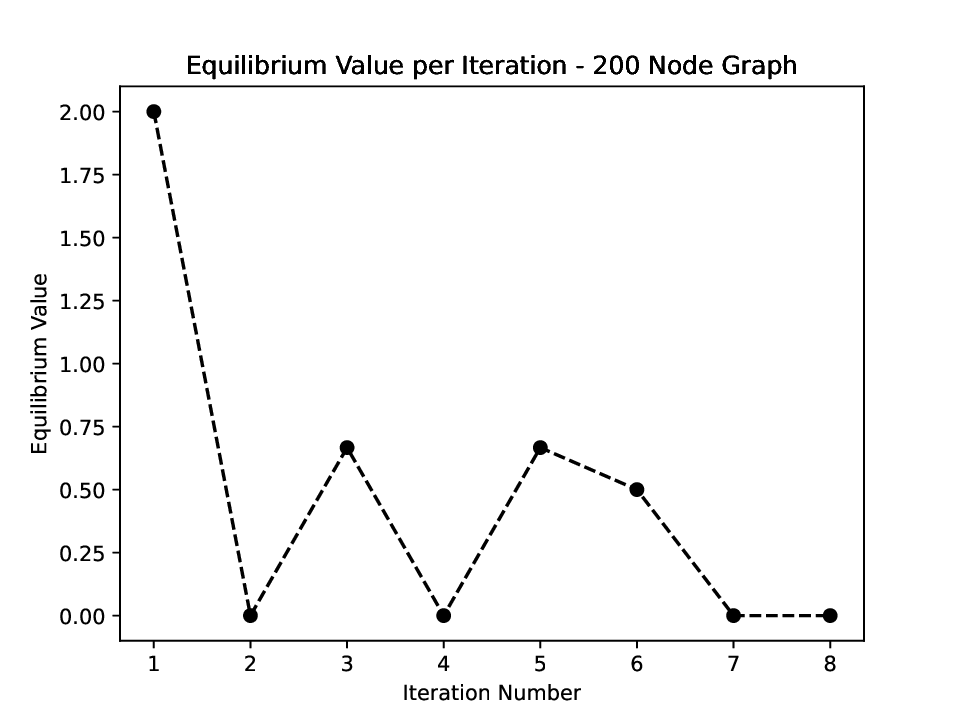}
        \caption{Sub-game Equilibrium Values}
        \label{fig:200-values}
    \end{subfigure}

    \caption{Algorithm execution for a single trial of the 200-vertex graph trial using the Patrolling defender policy. (a) The weight of each attacker heuristic in the sub-game equilibrium. (b) The sub-game equilibrium value at each iteration.}
    \label{fig:200-stats}
\end{figure}

We conclude our experiments with a case study on one configuration of the 200-vertex OSM graph (Figure~\ref{fig:osm200}).
In theory, this particular trial should be favorable to the Patrolling policy as the flag vertices are clustered close together, forming a tight patrol route. That said, the policy struggles against the attack in a full-visibility setting with an equilibrium of 2 flag captures out of a possible 2. The algorithm's execution is visualized in Figure~\ref{fig:200-stats}.

Figure~\ref{fig:200-mixed} visualizes how the attacker sub-game evolves over time. Initially, the Receding Horizon (green) heuristic is the dominant policy. However, as the Profile Oracle begins to adapt, we begin to see mixing between the Receding Horizon and Shortest Path (blue) policies starting by the third iteration. This continues until iteration 5, where the Shortest Path is subsumed by the Evasive Routing policy (pink). An interesting observation is that whenever a new attacker policy enters the sub-game, it is given the highest weight of any policy in that equilibrium. This could imply that the profiles fitted in each equilibrium are highly specialized to the attacker policies in that equilibrium, leading to poor performance against new additions to the sub-game.

\begin{figure}[htbp]
    \centering
        \begin{subfigure}[b]{0.42\textwidth}
        \centering

        \includegraphics[width=\textwidth]{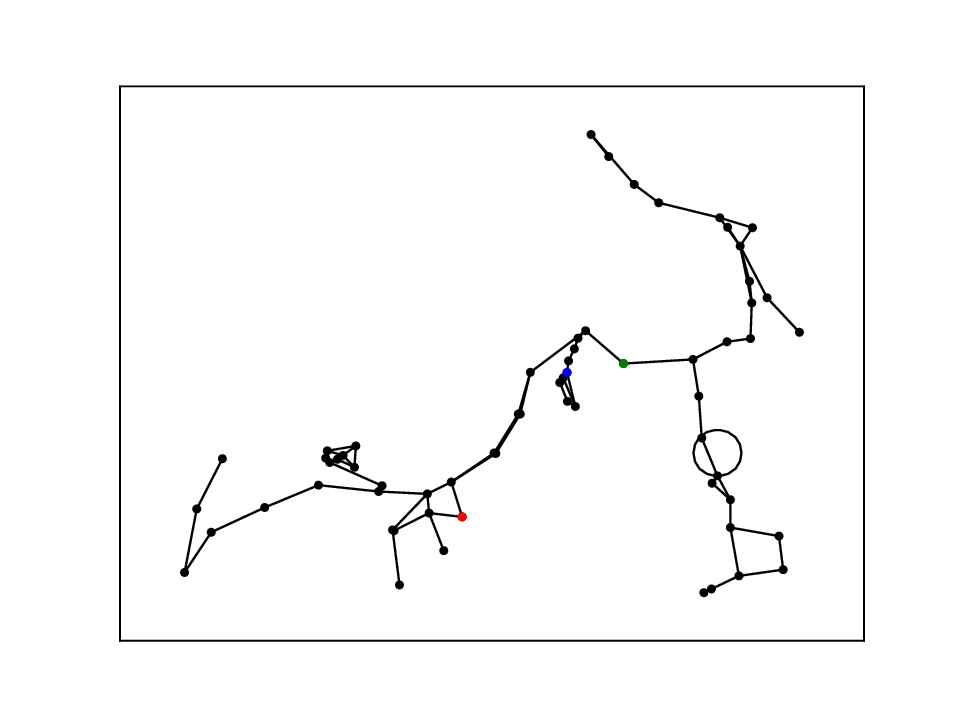}
        \caption{First defender graph}
        \label{fig:200-P1}
    \end{subfigure}
    \hfill 
    \begin{subfigure}[b]{0.42\textwidth}
        \centering
        \includegraphics[width=\textwidth]{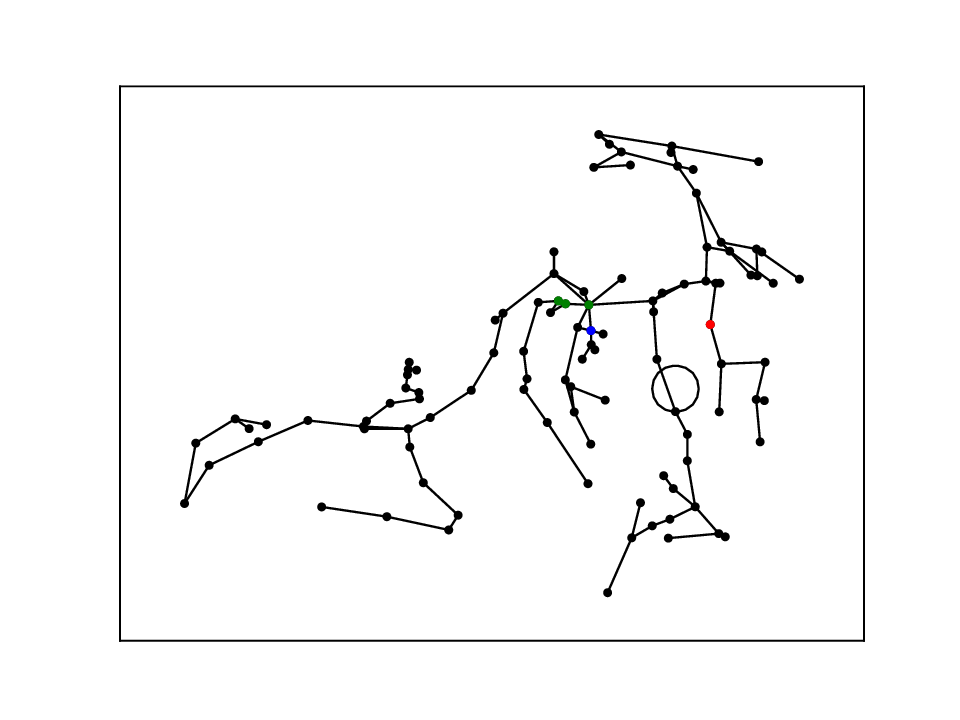}
        \caption{Second defender graph}
        \label{fig:200-P2}
    \end{subfigure}

    \caption{The final visibility profile generated on the 200-node OSM graph.}
    \label{fig:200-part}
    
\end{figure}

The expected number of captures at the sub-game equilibrium over the algorithm's execution is shown in Figure~\ref{fig:200-values}. In early iterations, we see that the number of captures fluctuates between positive values and zero. Specifically, we see a pattern of a new attacker policy being added to the subgame leading to a positive number of captures (iterations 1 and 3) and the profile adapting to get zero captures in the subsequent iteration (iterations 2 and 4). This pattern breaks in iteration 6, where the defense is not able to isolate a profile that can perfectly match the attacker's mixed strategy (i.e., zero flag captures). The profile generated at the end of iteration 7 does manage to achieve zero flag captures for the defense, representing a maximal improvement from the full-visibility setting. The final observable subgraphs for each player are shown in Figure~\ref{fig:200-part}.

\section{Conclusions and Future Directions}
\label{sec:conclusion}
We introduced a multi-agent zero-sum CtF game in which teams utilize decentralized heuristics rather than optimal policies. Each team takes a separate approach to mitigating its opponent's policies, with the attackers using heuristic diversity and the defense relying on altering its agents' observations. We proposed a Double Oracle algorithm for solving this game, where the Profile Oracle is approximated by a GNN. Our experiments show that our algorithm and approximate oracle can be effective in practice, and that altering agents' observations can elicit improved behaviors from a fixed policy.

A natural extension would be to allow both teams to reason over both a policy library and the space of visibility profiles. Furthermore, as observed in the 200-vertex case study, profiles generated as best responses tend to be highly specialized to the current attacker policies weighted in the sub-game. This tendency to overfit on seen policies remains a critical limitation of our approach. Thus, evaluating the transferability of these visibility profiles against policies outside of \(\Pi_{a}\) or developing methodologies to improve  their generalization is an important direction for future work.  Finally, expanding the formulation to allow for time-varying observation control and policy adaptation could provide an even greater range of behaviors to the defense than the static formulation considered here.

\begin{credits}
\subsubsection{\ackname} This work was sponsored by the Army Research Laboratory and was accomplished under Cooperative Agreement Number W911NF-17-2-0181. The authors would like to thank Dr. Amanda Prorok at the University of Cambridge for several helpful discussions regarding this work, Arijit Bhowmik at Michigan State University for contributing the Patrolling Defender and Receding Horizon attacker policies used in our experiments, and Hanqing Qi at George Mason University for his assistance in implementing the simulation environment.

\subsubsection{\discintname}
The authors have no competing interests to declare that are relevant to the content of this article.
\end{credits}


\bibliographystyle{splncs04}
\bibliography{refs}

\end{document}

%% file: tikz-images/algo-diagram.tex
\begin{tikzpicture}[
    >=stealth,
    node distance=1.5cm and 2cm,
    font=\small,
    box/.style={rectangle, draw=black, thick, rounded corners, align=center, minimum width=3cm, minimum height=1cm, fill=white},
    subbox/.style={rectangle, draw=black, thin, align=center, minimum width=2.5cm, minimum height=0.6cm, fill=gray!10, font=\footnotesize},
    dashedbox/.style={rectangle, draw=gray, thick, dashed, rounded corners, inner sep=10pt}
]

\node[box, fill=blue!5] (subgame) {Current Sub-game\\Matrix $U_k$};

\node[box, below=0.5cm of subgame] (nash) {LP Solver\\Nash Eq. $(y_k, z_k)$};

\node[box, below left=0.25cm and 0.5cm of nash] (profile_oracle) {Profile Oracle\\($p_{br} \in \mathcal{P}$)};
\node[box, below right=0.25cm and 0.5cm of nash] (attacker_oracle) {Attacker Oracle\\($\pi_{br} \in \Pi_a$)};

\node[subbox, below=0.8cm of profile_oracle] (actor_critic) {GNN Actor-Critic};
\node[subbox, below=0.3cm of actor_critic] (masking) {Autoregressive Masking};
\node[subbox, below=0.3cm of masking] (sim) {Simulation \& Reward};

\node[dashedbox, fit=(actor_critic) (masking) (sim)] (rl_box) {};
\node[anchor=south west, font=\footnotesize\itshape, text=gray] at (rl_box.north west) {DRL Approximation};

\node[subbox, below=1.5cm of attacker_oracle] (exact) {Exact evaluation over $\Pi_a$};

\draw[->, thick] (subgame) -- (nash);
\draw[->, thick] (nash) -| node[above, pos=0.5] {Defender Strategy $y_k$} (attacker_oracle);
\draw[->, thick] (nash) -| node[above, pos=0.5] {Attacker Strategy $z_k$} (profile_oracle);

\draw[->] (profile_oracle) -- (actor_critic);
\draw[->] (actor_critic) -- (masking);
\draw[->] (masking) -- (sim);

\draw[->] (attacker_oracle) -- (exact);

\draw[->, thick, rounded corners] (rl_box.west) -| ([xshift=-1.5cm]rl_box.west) |- node[above, pos=0.75] {Add $p_{br}$ to $Y$} (subgame.west);
\draw[->, thick, rounded corners] (exact.east) -| ([xshift=1.5cm]exact.east) |- node[above, pos=0.75] {Add $\pi_{br}$ to $Z$} (subgame.east);

\end{tikzpicture}

%% file: tikz-images/gridworld.tex
    \begin{tikzpicture}[
    level distance ={15mm},
    root/.style={draw=black, circle, fill=blue, text=black, minimum size = 2mm},
    gen/.style ={draw=black,  circle, fill=black, minimum size = 2mm},
    perim/.style={draw=black, circle, fill=red, text=black, minimum size = 2mm},
    int/.style={draw=green, circle, fill=green, text=black, minimum size = 2mm},
    dummy/.style={circle}, 
    ]
    % \draw[help lines] (-3,-4) grid (3,0);
    \useasboundingbox (-4, -5) rectangle (4, 1);
    \node[perim] (0) at (-3,-4) {};
    \node[gen] (1) at (-1,-4) {};
    \node[gen] (2) at (1,-4) {};
    \node[gen] (3) at (3,-4) {};
    % \node[gen, opacity=0.3] (4) at (-3,-2) {};
    % \node[gen, opacity=0.3] (5) at (-1,-2) {};
    % \node[gen, opacity=0.3] (6) at (1,-2) {};
    \node[gen] (4) at (-3,-2) {};
    \node[gen] (5) at (-1,-2) {};
    \node[gen] (6) at (1,-2) {};
    \node[gen] (7) at (3,-2) {};
    \node[root] (8) at (-3,0) {};
    \node[gen] (9) at (-1,0) {};
    \node[gen] (10) at (1,0) {};
    \node[int] (11) at (3,0) {};
    \draw  (0) -- (1);
    \draw  (1) -- (2);
    \draw  (2) -- (3);
    % dashed, gray
    \draw  (4) -- (5);
    \draw  (5) -- (6);
    \draw  (6) -- (7);
    \draw  (8) -- (9);
    \draw  (9) -- (10);
    \draw  (10) -- (11);
    \draw  (0) -- (4);
    \draw  (4) -- (8);
    \draw  (1) -- (5);
    \draw  (5) -- (9);
    \draw (2) -- (6);
    \draw (6) -- (10);
    \draw  (3) -- (7);
    \draw  (7) -- (11);

\end{tikzpicture}

%% file: tikz-images/gw-base.tex
\begin{tikzpicture}[
    level distance ={15mm},
    root/.style={draw=black, circle, fill=blue, text=black, minimum size = 2mm},
    gen/.style ={draw=black,  circle, fill=black, minimum size = 2mm},
    perim/.style={draw=black, circle, fill=red, text=black, minimum size = 2mm},
    int/.style={draw=green, circle, fill=green, text=black, minimum size = 2mm},
    dummy/.style={circle}, 
    ]
    % \draw[help lines] (-3,-4) grid (3,0);
    \useasboundingbox (-4, -5) rectangle (4, 1);
    \node[perim] (0) at (-3,-4) {};
    \node[gen] (1) at (-1,-4) {};
    \node[gen] (2) at (1,-4) {};
    \node[gen] (3) at (3,-4) {};
    % \node[gen, opacity=0.3] (4) at (-3,-2) {};
    % \node[gen, opacity=0.3] (5) at (-1,-2) {};
    % \node[gen, opacity=0.3] (6) at (1,-2) {};
    \node[gen] (4) at (-3,-2) {};
    \node[gen] (5) at (-1,-2) {};
    \node[gen] (6) at (1,-2) {};
    \node[gen] (7) at (3,-2) {};
    \node[root] (8) at (-3,0) {};
    \node[gen] (9) at (-1,0) {};
    \node[gen] (10) at (1,0) {};
    \node[int] (11) at (3,0) {};
    \draw  (0) -- (1);
    \draw  (1) -- (2);
    \draw  (2) -- (3);
    % dashed, gray
    \draw  (4) -- (5);
    \draw  (5) -- (6);
    \draw  (6) -- (7);
    \draw  (8) -- (9);
    \draw  (9) -- (10);
    \draw  (10) -- (11);
    \draw  (0) -- (4);
    \draw  (4) -- (8);
    \draw  (1) -- (5);
    \draw  (5) -- (9);
    \draw (2) -- (6);
    \draw (6) -- (10);
    \draw  (3) -- (7);
    \draw  (7) -- (11);

    % \draw[->, red, ultra thick] () to [bend right] node[midway, above] {1} ();
    % \draw[->, blue, ultra thick] () to [bend right] node[midway, above] {1} ();
    \draw[->, red, ultra thick] (0) to [bend right] node[midway, below] {1} (1);
    \draw[->, red, ultra thick] (1) to [bend right] node[midway, below] {2} (2);
    \draw[->, red, ultra thick] (2) to [bend right] node[midway, below] {3} (3);
    \draw[->, red, ultra thick] (3) to [bend right] node[midway, right] {4} (7);
    \draw[->, red, ultra thick] (7) to [bend right] node[midway, right] {5} (11);

    \draw[->, blue, ultra thick] (8) to [bend left] node[midway, right] {1} (4);
    \draw[->, blue, ultra thick] (4) to [bend left] node[midway, right] {2} (0);
    \draw[->, blue, ultra thick] (0) to [bend left] node[midway, above] {3} (1);
    \draw[->, blue, ultra thick] (1) to [bend left] node[midway, above] {4} (2);
    \draw[->, blue, ultra thick] (2) to [bend left] node[midway, above] {5} (3);
    % \draw[->, red, ultra thick] (0) -- (1);
    % \draw[->, red, ultra thick] (1) -- (2);
    % \draw[->, red, ultra thick] (2) -- (3);
    % \draw[->, red, ultra thick] (3) -- (7);
    % \draw[->, red, ultra thick] (7) -- (11);
    % \draw[->, blue, ultra thick] (8) -- (9);
    % \draw[->, blue, ultra thick] (9) -- (10);
    % \draw[->, blue, ultra thick] (10) -- (11);
    % \draw[->, blue, ultra thick] (11) -- (7);
    % \draw[->, blue, ultra thick] (-3.5,0) -- (-3.5,-2);
    % \draw[->, blue, ultra thick] (-3.5,-2) -- (-3.5,-4);
    % \draw[->, blue, ultra thick] (-3,-4.5) -- (-1,-4.5);
    % \draw[->, blue, ultra thick] (-1,-4.5) -- (1,-4.5);
    % \draw[->, blue, ultra thick] (1,-4.5) -- (3,-4.5);

\end{tikzpicture}

%% file: tikz-images/gw-beats-ret.tex
    \begin{tikzpicture}[
    level distance ={15mm},
    root/.style={draw=black, circle, fill=blue, text=black, minimum size = 2mm},
    gen/.style ={draw=black,  circle, fill=black, minimum size = 2mm},
    perim/.style={draw=black, circle, fill=red, text=black, minimum size = 2mm},
    int/.style={draw=green, circle, fill=green, text=black, minimum size = 2mm},
    dummy/.style={circle}, 
    ]
    % \draw[help lines] (-3,-4) grid (3,0);
    \useasboundingbox (-4, -5) rectangle (4, 1);
    \node[perim] (0) at (-3,-4) {};
    \node[gen] (1) at (-1,-4) {};
    \node[gen] (2) at (1,-4) {};
    \node[gen] (3) at (3,-4) {};
    % \node[gen, opacity=0.3] (4) at (-3,-2) {};
    % \node[gen, opacity=0.3] (5) at (-1,-2) {};
    % \node[gen, opacity=0.3] (6) at (1,-2) {};
    \node[gen, opacity=0.3] (4) at (-3,-2) {};
    \node[gen] (5) at (-1,-2) {};
    \node[gen] (6) at (1,-2) {};
    \node[gen, opacity=0.3] (7) at (3,-2) {};
    \node[root] (8) at (-3,0) {};
    \node[gen] (9) at (-1,0) {};
    \node[gen] (10) at (1,0) {};
    \node[int] (11) at (3,0) {};
    \draw  (0) -- (1);
    \draw  (1) -- (2);
    \draw  (2) -- (3);
    % [dashed, gray]
    \draw[dashed, gray]  (4) -- (5);
    \draw  (5) -- (6);
    \draw[dashed, gray]  (6) -- (7);
    \draw  (8) -- (9);
    \draw  (9) -- (10);
    \draw  (10) -- (11);
    \draw[dashed, gray]  (0) -- (4);
    \draw[dashed, gray]  (4) -- (8);
    \draw  (1) -- (5);
    \draw  (5) -- (9);
    \draw (2) -- (6);
    \draw (6) -- (10);
    \draw[dashed, gray]  (3) -- (7);
    \draw[dashed, gray]  (7) -- (11);

    % \draw[->, red, ultra thick] () to [bend right] node[midway, above] {1} ();
    % \draw[->, blue, ultra thick] () to [bend right] node[midway, above] {1} ();
    \draw[->, red, ultra thick] (0) to [bend right] node[midway, below] {1} (1);
    \draw[->, red, ultra thick] (1) to [bend right] node[midway, above] {2} (0);
    \draw[->, red, ultra thick] (0) to [bend left] node[midway, left] {3} (4);
    \draw[->, red, ultra thick] (4) to [bend left] node[midway, left] {4} (8);
    \draw[->, red, ultra thick] (8) to [bend left] node[midway, above] {5} (9);

    \draw[->, blue, ultra thick] (8) to [bend right] node[midway, below] {1} (9);
    \draw[->, blue, ultra thick] (9) to [bend left] node[midway, right] {2} (5);
    \draw[->, blue, ultra thick] (5) to [bend left] node[midway, right] {3} (1);
    \draw[->, blue, ultra thick] (1) to [loop below] node[midway, below] {4} (1);
    \draw[->, blue, ultra thick] (1) to [bend left] node[midway, left] {5} (5);

\end{tikzpicture}

%% file: tikz-images/gw-final.tex
    \begin{tikzpicture}[
    level distance ={15mm},
    root/.style={draw=black, circle, fill=blue, text=black, minimum size = 2mm},
    gen/.style ={draw=black,  circle, fill=black, minimum size = 2mm},
    perim/.style={draw=black, circle, fill=red, text=black, minimum size = 2mm},
    int/.style={draw=green, circle, fill=green, text=black, minimum size = 2mm},
    dummy/.style={circle}, 
    ]
    % \draw[help lines] (-3,-4) grid (3,0);
    \useasboundingbox (-4, -5) rectangle (4, 1);
    \node[perim] (0) at (-3,-4) {};
    \node[gen] (1) at (-1,-4) {};
    \node[gen, opacity=0.3] (2) at (1,-4) {};
    \node[gen] (3) at (3,-4) {};
    % \node[gen, opacity=0.3] (4) at (-3,-2) {};
    % \node[gen, opacity=0.3] (5) at (-1,-2) {};
    % \node[gen, opacity=0.3] (6) at (1,-2) {};
    \node[gen, opacity=0.3] (4) at (-3,-2) {};
    \node[gen] (5) at (-1,-2) {};
    \node[gen] (6) at (1,-2) {};
    \node[gen] (7) at (3,-2) {};
    \node[root] (8) at (-3,0) {};
    \node[gen] (9) at (-1,0) {};
    \node[gen, opacity=0.3] (10) at (1,0) {};
    \node[int] (11) at (3,0) {};
    \draw  (0) -- (1);
    \draw[dashed, gray]  (1) -- (2);
    \draw[dashed, gray]  (2) -- (3);
    % [dashed, gray]
    \draw[dashed, gray]  (4) -- (5);
    \draw  (5) -- (6);
    \draw  (6) -- (7);
    \draw  (8) -- (9);
    \draw[dashed, gray]  (9) -- (10);
    \draw[dashed, gray]  (10) -- (11);
    \draw[dashed, gray] (0) -- (4);
    \draw[dashed, gray]  (4) -- (8);
    \draw  (1) -- (5);
    \draw  (5) -- (9);
    \draw[dashed, gray] (2) -- (6);
    \draw[dashed, gray] (6) -- (10);
    \draw  (3) -- (7);
    \draw  (7) -- (11);

    \draw[->, red, ultra thick] (0) to [bend right] node[midway, below] {1} (1);
    \draw[->, red, ultra thick] (1) to [bend right] node[midway, below] {2} (2);
    \draw[->, red, ultra thick] (2) to [bend right] node[midway, below] {3} (3);
    \draw[->, red, ultra thick] (3) to [bend right] node[midway, right] {4} (7);
    % \draw[->, red, ultra thick] (8) to [bend right] node[midway, above] {5} (9);

    \draw[->, blue, ultra thick] (8) to [bend left] node[midway, above] {1} (9);
    \draw[->, blue, ultra thick] (9) to [bend left] node[midway, right] {2} (5);
    \draw[->, blue, ultra thick] (5) to [loop below] node[midway, below] {3} (5);
    \draw[->, blue, ultra thick] (5) to [bend left] node[midway, above] {4} (6);
    % \draw[->, blue, ultra thick] (1) to [bend left] node[midway, left] {5} (5);

\end{tikzpicture}